\documentclass[%
 reprint,
 amsmath,amssymb,
 aps,
]{revtex4-2}

\usepackage{graphicx}
\usepackage{dcolumn}
\usepackage{bm}
\usepackage{hyperref}
\usepackage[mathlines]{lineno}

\begin{document}

\preprint{APS/123-QED}
\title{Electrostatic tuning of bilayer graphene edge modes}

\author{%
Hira Ali}%

 
\author{%
Llorenç Serra}%
\email{hira@ifisc.uib-csic.es (H.A)}
 \email{llorens.serra@uib.es (L.S)}
\affiliation{%
Institute for Cross-Disciplinary Physics and Complex Systems IFISC (CSIC-UIB), E-07122 Palma, Spain\\
 Physics Department, University of the Balearic Islands, E-07122 Palma, Spain\\
 }%



\begin{abstract}
 We study the effect of a local potential shift induced by a side electrode 
on the edge modes at the boundary between gapped and ungapped bilayer graphene. A potential shift close to the gapped-ungapped boundary
causes the emergence of 
unprotected edge modes, propagating in both directions along the boundary. These counterpropagating edge modes allow edge backscattering, as opposed to the case of valley-momentum-locked  edge modes. 
We then calculate
the conductance of a bilayer graphene wire in presence of finger-gate electrodes, finding strong asymmetries with energy inversion and deviations from conductance quantization that can be understood with the gate-induced unprotected edge modes.

\end{abstract}

\maketitle


\section{\label{Intro} Introduction  }

\begin{figure}[t]
\begin{center}    
\includegraphics[width=0.50\textwidth, trim = 1.7cm 10.5cm 3.2cm 2.5cm, clip]{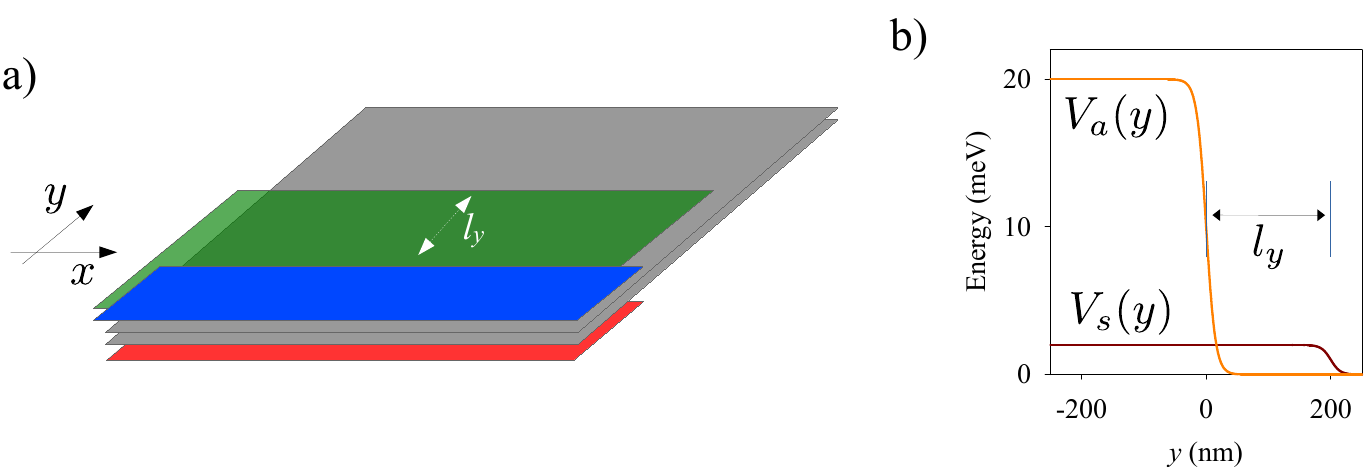}
\end{center}
\caption{
a) Sketch showing the two bilayer graphene planes (gray), the pair of electrodes for asymmetric potential $V_a$ (red and blue) and the electrode for symmetric potential $V_s$ (green). A $y$-displacement of the symmetric and asymmetric electrodes is indicated by  $l_y$.
b) Model symmetric $V_s(y)$ and asymmetric $V_a(y)$ potentials with a displacement $l_y=200$ nm, $V^0_s=2$ meV, $V^0_a=20$ meV. 
}
\label{Fig1}
\end{figure}

Bilayer graphene (BLG) allows a remarkable mechanism of electronic confinement by
tuning the energy gap with electrostatic gates on the sides of the two graphene layers.\cite{Mcan13, Zhang13, rozhkov16, Over18, Kraf18, Eich18, Kurzmann19, Banszerus20, Banszerus21, Ban23}
Indeed, an interlayer electric field opens a gap in the spectrum, thus favouring electronic confinement to those regions with a vanishing (or small) 
interlayer field. 
Electrodes of carefully chosen shapes, designed  with lithographic techniques, allow different types of BLG nanostructures such as open semi-infinite edges, quasi-1D wires (electron guides),
and fully closed loops, rings and dots. For instance, the blue/red 
electrodes in 
Fig.\ \ref{Fig1}a create an open BLG edge at $y=0$, separating 
two half planes, gapped and ungapped, for electronic motion.

Graphene nanostructures can also be made with etching techniques,
removing parts of the graphene system,
as opposed to the above mentioned electrostatic
confinement of BLG. With etching, however, the specific atomic arrangement at the borders as well as the presence of undesired edge roughnesses or imperfections is usually relevant and methods 
to reduce or minimize them are generally 
desired.\cite{Men12,Cle19,Jin21} 

The tuning of the electric gap in BLG using electric fields was demonstrated in early magnetotransport experiments with bulk BLG \cite{Cas07}. These were followed by a very intense research activity, 
as summarized, e.g., in the field reviews 
Refs.\ \cite{Mcan13,rozhkov16}.
More recently, experiments on electrostatic confinement in
BLG nanostructures have been reported for
dots \cite{Eich18,Kurzmann19,Banszerus20,Banszerus21,Ban23}
and 1D edges
\cite{Lon15,Men15,Li16,Cle19,Chen20}.
The hallmark of the latter are the observation of conductance quantization in quantum transport experiments.

Ungapped BLG hosts bulk propagating electronic modes for any energy, with characteristic  2D wave numbers,
$(k,q)$  in $(x,y)$ directions. States in  translationally invariant edges 
or wires in only one direction ($x$) have a 1D wave number ($k$);
while closed loops and dots 
possess a fully discrete electronic spectrum. In Ref.\ \cite{Ryu22} it was shown 
that an open electrostatic edge in BLG
is able to bind an
edge mode with a characteristic valley-momentum locking; i.e., with opposite valleys propagating in opposite directions along the edge. Remarkably, the wave number $k$  of this mode 
separates from the continuum band of bulk ungapped modes and yields
characteristic transport signatures in BLG junctions.\cite{Ryu22}

In this work we further investigate the properties of electrostatic 
edge modes in BLG. 
In particular, we focus on the effect of a potential shift
as induced by an additional side electrode. We consider an electrostatic edge defined by two side gates (red and blue in Fig.\ref{Fig1}a), and an additional gate creating the potential shift (green in Fig. \ref{Fig1}a). We found that a lateral shift 
of the electrodes by a small distance $l_y$ has a very relevant effect. It causes additional edge modes in the stripe of width $l_y$, running in both directions along the edge. 
Therefore, they allow backscattering mediated by these edge modes alone, without the need to couple with bulk modes. In presence of disorder, or 
other inhomogeneities, an additional electrode will then strongly affect the conductance 
along the edge. Besides, the electrode also causes energy-inversion asymmetry, with different conductances for positive and negative energies
(with zero energy being the Dirac-point reference energy).

Subsequently, we use the results on the open edge to understand the effect of finger gate electrodes (FGE) across a BLG wire or guide. We consider the cases of an extended FGE covering both wire edges, or shorter FGE's affecting one or the two edges 
of the wire. We predict conspicuous energy asymmetries and conductance deviations from quantization that can be explained with the FGE induced edge modes. Therefore, similarly to the case of semiconductor wires,\cite{Ihn} FGE's are a practical way to manipulate electronic transport in BLG electrostatic wires.

\section{ Theoretical Model }

Our analysis is based on a low-energy Hamiltonian for BLG in presence 
of electrostatic potentials.\cite{Mcan13}
We consider two types of potentials: a
{\em symmetric} potential $V_s$, equal on the two layers, and an 
{\em asymmetric} potential $\pm V_a$, with opposite signs on the two 
layers. In this work we consider parameterized model functions for 
both potentials, as shown in Fig. \ref{Fig1}b for the case of an open edge.
These functions read
\begin{equation}
 V_{s/a}(y) =  \frac{V^{0}_{s/a}}{1+e^{(y-y_{s/a})/s}}\; ,
\label{eq0}
 \end{equation}
where parameters $V^0_{s/a}$ and $y_{s/a}$ are 
the asymptotic 
value and position of the border for the symmetric/asymmetric
potential. Parameter $s$ is 
a small distance representing the smoothness of the potential steps. 
Examples of our model potentials can be seen in Fig.\ref{Fig1}b.

The low-energy effective Hamiltonian 
we will use in this work
is built on an underlying 
tight-binding atomistic description for BLG.  The electronic band struture of unbiased bulk BLG
is characterized by gap closings at the six Dirac points
in reciprocal space, three of them corresponding to the valley $K_+$ and the other three to valley $K_-$.
Near those Dirac points,
an expansion to 
the leading terms in electronic momenta 
yields an effective multiband continuum Hamiltonian.
We  refer the reader to Ref. \cite{Mcan13} for details on the mathematical derivations and only
stress here that we restrict to 
graphene layers in AB Bernal stacking.
Adding the model potentials of the type (\ref{eq0})  
to the resulting effective Hamiltonian describes the specific confinement mechanisms due to the electrostatic gates of this work. The potential  
difference $2V_a$ between the two graphene layers opens an energy gap in the low-energy scale around the Dirac points
which is modulated in space by a position dependent potential.

The BLG low-energy Hamiltonian reads \cite{Mcan13}
\begin{eqnarray}
 H &=& v_F p_x \tau_z \sigma_x
 + v_F\, 
p_y \sigma_y 
 + \frac{t}{2}\, \left(\,\lambda_x \sigma_x +\lambda_y\sigma_y\,\right)
 \nonumber\\
 &+& V_s(x,y)+ V_a(x,y)\, \lambda_z\; ,
\label{eq1}
 \end{eqnarray}
with the Fermi velocity 
$\hbar v_F= 660\, {\rm meV}\,{\rm nm}$  and the interlayer coupling $t=380\,{\rm meV}$. In Eq.\ (\ref{eq1}), $\sigma_{x,y,z}$, $\tau_{x,y,z}$
and $\lambda_{x,y,z}$ are sets of Pauli matrices for sublattice, valley and layer
degrees of freedom, respectively. This Hamiltonian is valley diagonal and it has been used 
to study quantum states in a variety of BLG nanostructures.
As mentioned in Sec. \ref{Intro}, the use of position-dependent potentials $V_s(x,y)$ and $V_a(x,y)$ allow  modeling the 
effect of 
potential gates that create 
electrostatic borders. 
Notice that Eq. (\ref{eq1}) is for general inhomogenous potentials  $V_s$ and $V_a$ depending on both coordinates $(x,y)$.  See below, however, for the restricted cases 
considered in this work of potentials  which are
uniform along $x$ or piecewise-uniform along $x$, with each uniform section  of the type given in 
Eq. (\ref{eq0}).

We also stress that for the case of sign-changing
$V_a(x,y)$'s,  Hamiltonian (\ref{eq1}) predicts the emergence of {\em topological} modes 
near the sign-change border.
The spectrum becomes gapless in presence of these modes since
their energies $E(k)$ cross from the negative to the positive energy sectors.
These modes also show a 
characteristic valley-momentum locking and protection from bulk modes by an energy gap. 
\cite{Pereira07,Recher09,Zarenia09,Pereira09,Zarenia10,Zarenia10b,daCosta14}
From a formal Condensed Matter topology approach, it has been pointed out
that specific invariants for each valley $N_\tau=\pm1$ can be approximately 
defined in BLG.\cite{Li10,Zhang13} However, it has also been stressed
that the bulk-boundary correspondence between those invariants 
and the edge modes is not general and may depend on the 
specific type of interface, such as in BLG-vacuum or BLG-BLG.\cite{Li10}
The latter type
corresponding to the 
electrostatic boundaries considered in this work. 

\begin{figure}[t]
\begin{center}
\includegraphics[width=0.50\textwidth, trim = 1cm 10.cm 1cm 2cm, clip]{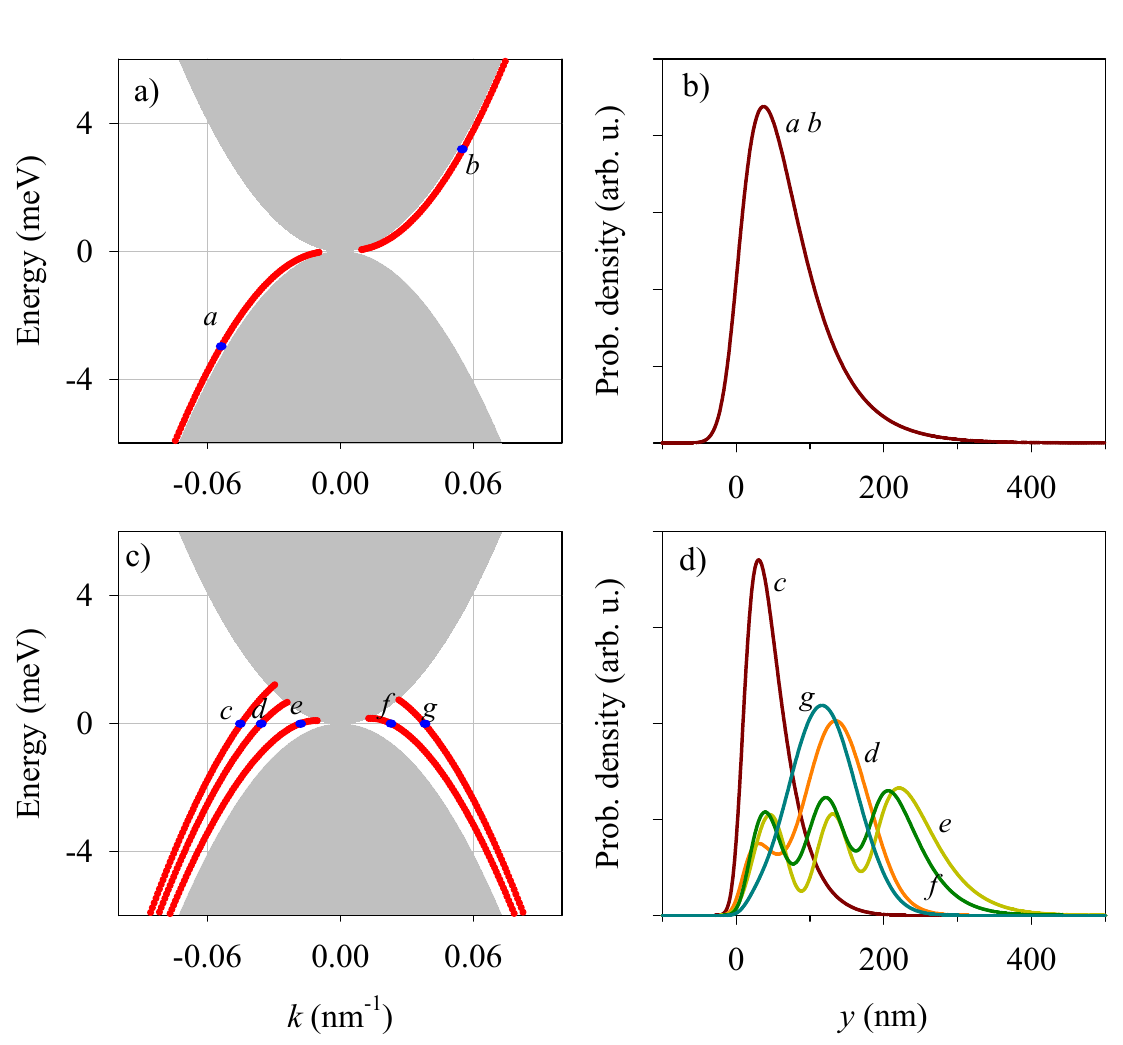}
\end{center}
\caption{
a) Eigenenergies for the (open) single edge with $V_s=0$. The gray region is the bulk continuum while the red line is the discrete branch of edge states. 
b) Spatial probability distributions for two selected wave numbers $k$
indicated in panel a) by the corresponding labels.
c,d) Similar results to a,b) but with $V_s=2$ meV and $l_y=200$ nm. 
Other parameters: $V_a=20$ meV, $s=7.5$ nm. 
}
\label{Fig2}
\end{figure}

Below, we will discuss a) the eigenstates of the Hamiltonian 
(\ref{eq1}) for fully translational invariant BLG systems
with both potentials $V_s(y)$ and $V_a(y)$;
and b) the conductance
through junctions of different BLG sections
described by $V_s(x,y)$ and $V_a(x,y)$ having a piecewise-constant dependence on $x$.
They model the effect  of a central FGE on a quantum wire (see device sketches in Figs.\ \ref{Fig3}-\ref{Fig5}).
In all cases our resolution method is based on a combination of spatial grid discretization and multiple component wave functions using complex-band-structure theory.    
More details of the method can be found in 
Sec.\ \ref{method}.

\section{ Results and Discussion }

\subsubsection{ Single Edge}

Figure \ref{Fig2} shows the electron eigenenergies for the open BLG edge
sketched in Fig. \ref{Fig1}. The gray region in Fig. \ref{Fig2}a is 
the continuum for bulk modes, given by the condition 
\begin{equation}
 |k| < 
 \frac{1}{\hbar v_F}\, \sqrt{ |E|\, \left(\, |E| + t\, \right)}\; .
\label{eq3}
\end{equation}
See App.\ \ref{bulk} for a derivation of this momentum restriction
for bulk propagating states.
The red line in Fig.\ \ref{Fig2}a shows the edge mode in absence of symmetric potential $V_s=0$. This mode 
spatially
decays with the distance
to the boundary (Fig.\ \ref{Fig2}b) and it is characterized by valley-momentum locking; reversed valleys propagating in reversed directions
in a similar way to the quantum spin Hall effect but replacing  spin
with valley.

The edge mode becomes damped when it overlaps with the continuum
of bulk BLG modes, indicated in gray colour in Fig.\ \ref{Fig2}a. In this case the localized edge mode decays into bulk modes with the same $E$ and $k$, thus flying away from the edge.
In Fig.\ \ref{Fig2}a this overlap occurs in the region of vanishing $E$ and $k$ 
and, technically, it is not easily resolved by the numerical calculation.

\begin{figure}[t]
\begin{center}
\includegraphics[width=0.55\textwidth, trim = 5.5cm 2.5cm 8cm 2cm, clip]{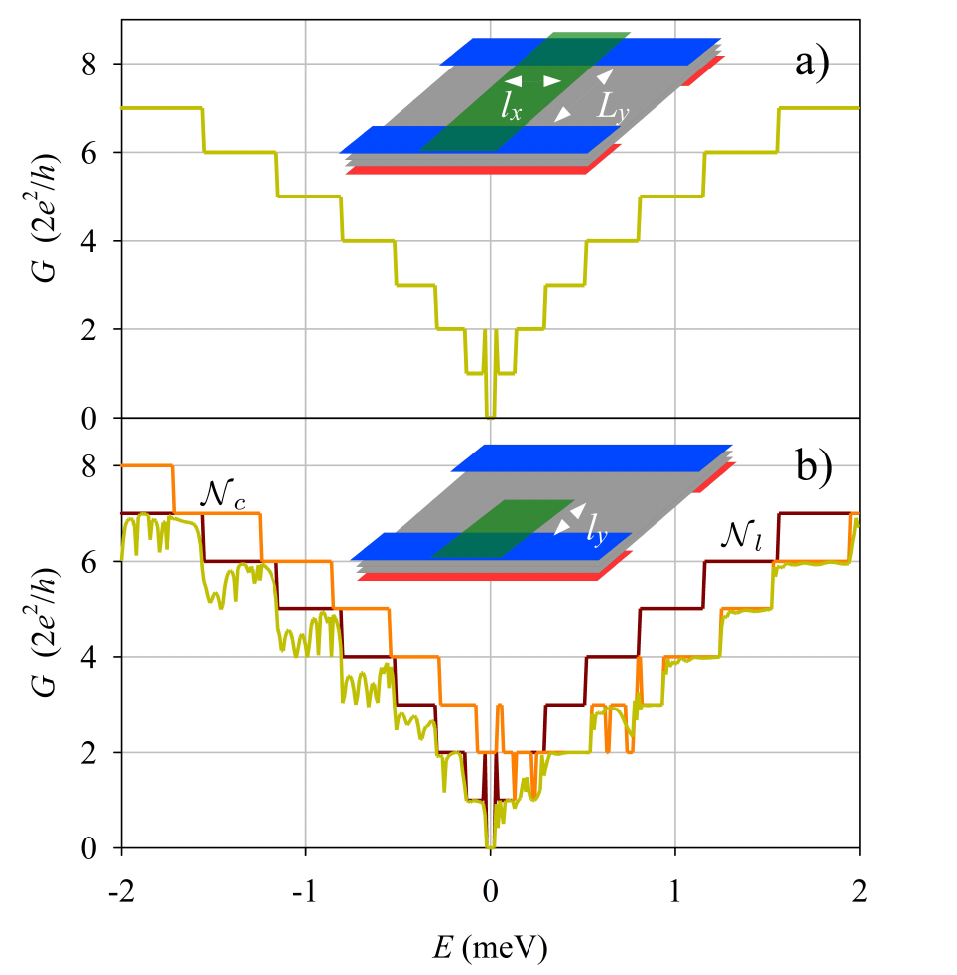}
\end{center}
\caption{a) Conductance of a quantum wire in presence of a FGE  across all the wire (a), and with a FGE covering only one edge (b). The sketch insets show the corresponding systems. The number of active modes in the asymptotic leads ${\cal N}_l$ and center ${\cal N}_c$ are also shown. Parameters: $L_y=600$ nm, $l_y=200$ nm, $l_x=1$ $\mu$m, $V_s^0=0.5$ meV.
}
\label{Fig3}
\end{figure}

\begin{figure}[t]
\begin{center}
\includegraphics[width=0.55\textwidth, trim = 5.5cm 2.5cm 8cm 2cm, clip]{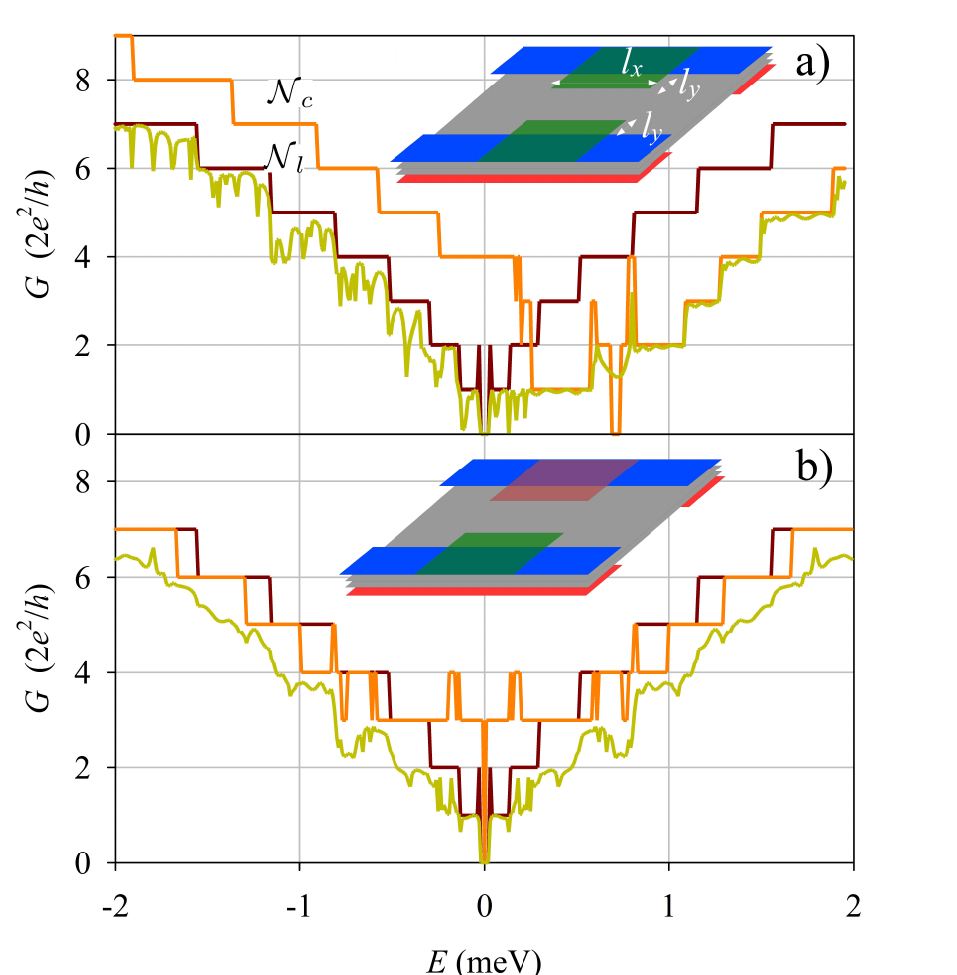}
\end{center}
\caption{Similar to Fig.\ \ref{Fig3} but with two FGE's with the same $V^0_s$ (a), and with opposite $V^0_s$ (b).
}
\label{Fig4}
\end{figure}

The modification induced by a potential shift of an additional gate
with $l_y=200$  nm 
is shown in Figs.\ \ref{Fig2}cd. 
The discrete branch of states of Fig.\ \ref{Fig2}a is now shifted upwards in energy, merging with the 
continuum for energies beyond a given maximum value. In addition, new branches of modes emerge at low and negative energies that are localized to the region of width $l_y$ near the boundary. These modes propagate in both directions, as seen from the positive and negative slopes
of the energy branches. The corresponding probability densities show  substantial overlaps (Fig.\ \ref{Fig2}d), suggesting the 
possibility of backscattering mediated by these edge modes in presence of inhomogeneities along the edge. Most remarkably, the additional side gate (and potential shift $V_s$) yields energy-inversion asymmetry
in Fig.\ \ref{Fig2}c, with the presence of edge-mode branches only in the lower part of the energy diagram. 
We stress that the shift $l_y$ in Fig.\ \ref{Fig1}a is essential for the emergence of
additional branches of edge states, as well as for the energy asymmetry 
of the spectra.

\subsubsection{Quantum Wire Junctions}

\begin{figure}[t]
\begin{center}
\includegraphics[width=0.55\textwidth, trim = 5.5cm 2.cm 8cm 2cm, clip]{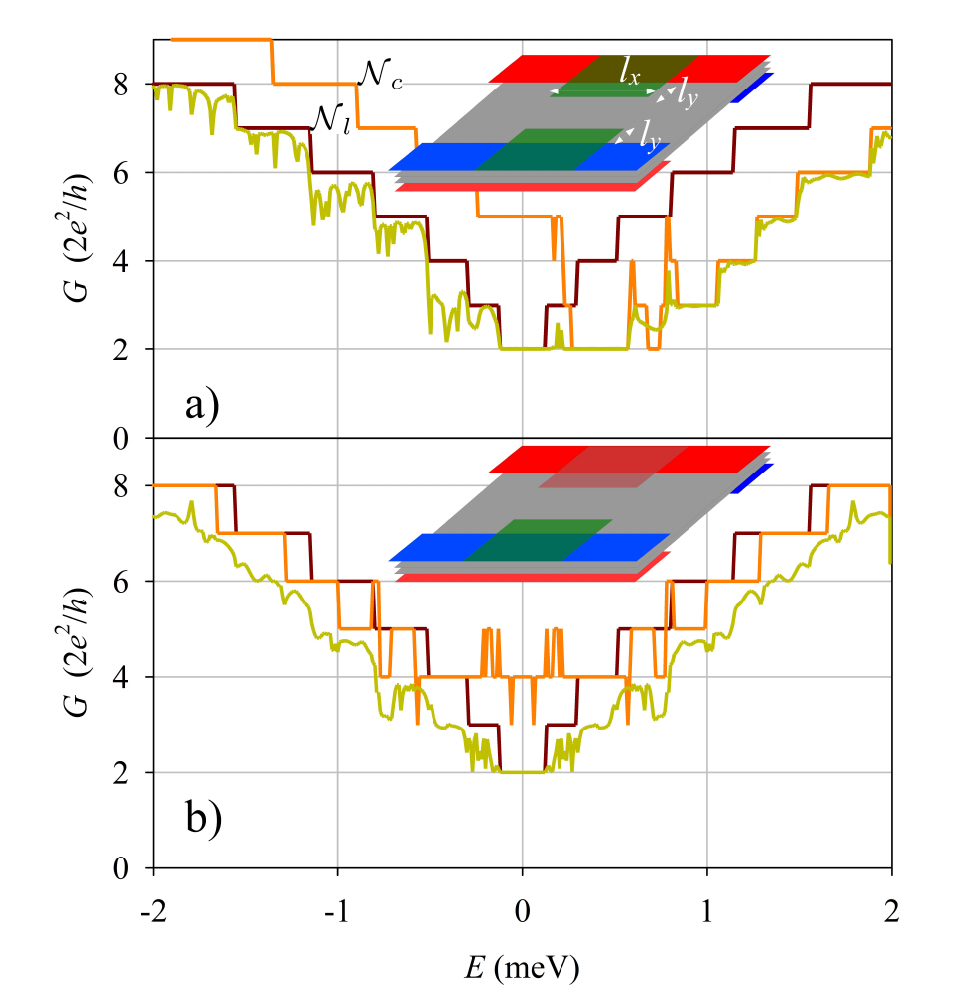}
\end{center}
\caption{ Similar to Figs.\ \ref{Fig3} and \ref{Fig4} but for a quantum wire with topological confinement; i.e., with reversed blue and red 
electrodes on the two wire edges, as shown in the inset sketches.}
\label{Fig5}
\end{figure}

 Having analyzed the gate-induced modifications in the open edge, we consider next the role of a FGE on an electrostatic quantum wire
of width $L_y$. More specifically, we calculate the total left-to-right
transmission $T$ (with conductance $G=T e^2/h$) using the complex-band-structure method for the double junction system sketched in Figs.\ \ref{Fig3}-\ref{Fig5}. Firstly, Fig.\ \ref{Fig3}a shows that a FGE covering all the wire has a negligible effect on the wire conductance: 
transmission is perfect and the coductance simply reproduces the 
staircase function of the number of active modes. 
On the contrary, a FGE covering only one edge of the quantum wire 
(Fig.\ \ref{Fig3}b) yields 
relevant modifications. $G$ deviates from the  plateau values, with conspicuous minima for energies $E<V_s^0$. There is also a clear asymmetry 
with respect to energy inversion in Fig.\ \ref{Fig3}b. For $E>V_s^0$ 
the conductance is almost perfectly quantized, while for $E<V_s^0$ it shows the mentioned deviations. 

The conductance non-quantization and asymmetry of Fig.\ \ref{Fig3}b
can be understood as effects of the edge modes induced by the 
FGE, as discussed above. Quasi bound states, allowed by edge mode backscatterings at the interfaces, lead to conductance dips for  
specific (resonant) energies. This mechanism is only present 
for $E<V_s^0$, thereby explaining the asymmetry in conductance.

The case of two FGE's, one on each edge of the wire, is presented in Fig.\ \ref{Fig4}. We studied this configuration using the same shift $V_s^0$
on the two FGE's (Fig.\ \ref{Fig4}a), and 
with opposite signs of the shift $\pm V_s^0$ on the two FGE's (Fig.\ \ref{Fig4}b). The case of identical shifts is very similar to the preceding case with just a single FGE (Fig.\ \ref{Fig3}b). However, with opposite signs the results change markedly; the conductance becoming again symmetric with energy inversion and the deviations from quantization are enhanced.

\begin{table}[t]
\begin{center}
\begin{tabular}{c|c|c}
\hline \hline
Confinement & FGE1 x FGE2 & conductance \\
\hline
Trivial    & +  & A \\
           & -  & S \\
\hline
           Topological & +   & A \\
           & -  & S \\          
\hline\hline
\end{tabular}
\end{center}
\caption{
Symmetric/antisymmetric (S/A) character of the conductances in Figs.\ \ref{Fig4} and \ref{Fig5}. 
The column FGE1xFGE2 indicates the sign product of the FGE potentials covering the two wire edges.
}
\label{tab1}
\end{table}

As a final case, we consider a topological inversion in the 
asymmetric potential $V_a$,\cite{Mar08,Zarenia11,xavier10,Ben21,Ben21b}
with the red/blue electrodes being reverted on the two edges of the quantum wire (Fig.\ \ref{Fig5}) and with two
FGE's. The results in 
Fig.\ \ref{Fig5}ab
are very similar to those in
Fig.\ \ref{Fig4}ab but with a notable difference near zero energy.
Namely, in the topological cases the conductance is perfectly quantized
to $4e^2/h$ in a small energy plateau around zero, 
while it vanishes in Figs.\ \ref{Fig3} and \ref{Fig4}.
This is explained by the gapless character of the topological wire, 
which hosts two valley-momentum-locked branches crossing zero energy. 
On the contrary, the nontopological (trivial) confinement in a finite 
$L_y$  wire 
is always characterized by a zero energy gap due to the finite size.

The energy inversion symmetries of the different configurations of FGE electrodes considered in
Figs.\ \ref{Fig4} and \ref{Fig5} 
is summarized in 
Table \ref{tab1}. Notice that this
symmetry is fixed by the product of the signs of FGE potentials on opposite edges of the wire, irrespectively of the trivial or topological character of the wire confinement.
This result illustrates how conductance measurements could be used to observe tuning of the edge modes using FGE's.

\subsubsection{ Further discussion}

All results presented above are for a single valley, $K_+$. The corresponding results for the reversed valley $K_-$ can be inferred simply reverting $k\to-k$ in Fig.\ \ref{Fig2}ab for a single edge, while the results remain invariant in the cases of wires with FGE's of Figs.\ \ref{Fig3}-\ref{Fig5}. Therefore, we do not find any valley polarization induced by a FGE in the quantum wire.  

An important underlying aspect, however beyond our present analysis, is  
the role of random imperfections and disorder in the device. 
While it can be reasonable to assume that BLG is relatively free of such disorder effects, the additional processing required for the electrostatic electrodes could introduce such random disorder. Therefore, this is a relevant aspect to consider in the future. 
We may expect, however, that the conductance asymmetry and the nonquantization  induced by FGE's in a quantum wire would be enhanced in presence of random disorder.   

\section{Conclusions}

We have studied the role of an electrode creating a potential shift near an electrostatic BLG edge. We found that in presence of a displacement $l_y$ between the BLG edge and the additional electrode, new modes emerge 
near the edge,
in the region of width $l_y$,  that propagate in both 
directions. Furthermore, the valley-momentum-locked branch of the 
single edge 
is shifted in energy by the electrode and the spectrum becomes asymmetric with energy inversion.

We also investigated the more practical case of a BLG quantum wire in 
presence of transverse FGE's. If a FGE is covering the two edges of a quantum wire, the system's conductance is almost unchanged and remains nearly perfectly quantized. However, if the FGE covers only one 
edge, or there are two different FGE's covering the two edges, then the conductance displays strong non quantizations and asymmetries with respect to energy inversion. These changes are in good agreement with the
modifications expected from the single-edge spectrum 
in presence 
and a displaced electrode. 
The energy inversion symmetry in the quantum wire is restored with two FGE's having opposite potential signs on the two edges. 
In summary, our work shows that FGE's can be a practical way to manipulate transport 
properties of BLG quantum wires by the electrostatic tuning the electronic modes at the 
wire edges.

\section{Methods}
\label{method}

We solved the eigen-problem with Hamiltonian (\ref{eq1}) using 
finite-difference discretization and matrix diagonalization routines.
With translational invariance, in the cases of a single edge and quantum wire, a matrix diagonalization for each 
(real) wave number $k$ yields the band structure $E_n(k)$ as well as it corresponding eigenstates.
An important aspect is the filtering of spurious modes emerging 
due to an artificial Fermion doubling of the physical eigenstates.
In practice the filtering is done by eliminating those states with  large oscillations in neigbouring grid points, such that spatially averaging on a small neigbourhood strongly modifies the wave function. We found this simple technique to be quite effective and robust.\cite{Ben21}

The transport problem for the junctions of piece wise homogenous sections in the transport direction ($x$) was solved using the complex-band-structure approach discussed in  Ref.\ \cite{Osca19}. Here, it is important to include complex wave numbers $k$ in order to describe evanescent-state behavior in the proximity of the junction interfaces. The wave-function matching at the junction interfaces is transformed into a large set of linear equations whose
solution determines the quantum transmissions $T_{kk'}$ and its corresponding Landauer conductance $G=\frac{e^2}{h}\sum_{kk'}{T_{kk'}}$. 

We refer to Ref.\ \cite{Osca19} for more details of the complex band structure approach and to Refs.\ \cite{Ben21,Ryu22} for its
specific application to BLG structures.


 \appendix
\section{Bulk continuum}

\label{bulk}

For constant $V_{s/a}$ potentials the eigenstates of the Hamiltonian (\ref{eq1}) are plane waves, with momenta $(k,q)$ along $(x,y)$,
\begin{equation}
\label{eq3A}
\Psi \equiv \Phi_{\eta_\sigma\eta_\tau\eta_\lambda}\, e^{i(kx+qy)}\; ,
\end{equation}
with $\eta_\sigma,\eta_\tau,\eta_\lambda=1,2$ indicating the different spinorial components of the wave function. Assuming a given real $k$, we can determine the corresponding $q$'s by 
transforming the eigenvalue equation as follows
\begin{equation}
H \Psi = E \Psi\quad\Rightarrow\quad
\sigma_y H \Psi = E \sigma_y \Psi\; .
\label{eq4A}
\end{equation}

The purpose of the above transformation is that 
Eq.\ (\ref{eq4A}) can be easily rewritten as 
an eigenvalue equation for $q$,
\begin{eqnarray}
\frac{1}{\hbar v_F}
\left[
\rule{0cm}{0.5cm}
E\,\sigma_y 
\right.
&+&
i \, \hbar v_F\,  k\, \sigma_z\tau_z 
+
\frac{t}{2}\left(
i\,\lambda_x\sigma_z -\lambda_y
\right) \nonumber\\
&-& V_s\,\sigma_y- \left. V_a\, \sigma_y \lambda_z 
\rule{0cm}{0.5cm}
\right]
\Phi
=
q\, \Phi\; .
\label{eq5A}
\end{eqnarray}
After some algebra the eigenvalues of Eq.\ (\ref{eq5A}), assuming
free BLG for which $V_{s/a}=0$, can be analytically determined 
with the diagonalization of an algebraic matrix.
The $q$ eigenvalues read
\begin{equation}
q=
\pm
\frac{1}{\hbar v_F}
\sqrt{
-\hbar^2 v_F^2 \, k^2 + |E| \left(\, |E| \pm t\, \right)
} \; .
\label{eq8A}
\end{equation}
Notice that Eq.\ (\ref{eq8A}) already proves the existence of a critical value
$k_c=
\sqrt{|E| \left(\, |E| + t\, \right)
}/\hbar v_F
$, as given in Eq.\ (\ref{eq3}). Indeed, the  
propagating-mode condition requires $q$ to be real which, from the square root in Eq.\ (\ref{eq8A}),
requires $|k|<k_c$.

\begin{acknowledgments}
We acknowledge support from Grant No.\ PDR2020-12 
funded by GOIB; and from Grant
No.\ MDM2017-0711
and Grant
No.\ PID2020-117347GB-I00
funded by MCIN/AEI/10.13039/501100011033.
H.A. was supported by GOIB program ``SOIB Recerca i Innovaci\'o'' 

\end{acknowledgments}

 \appendix



\bibliography{Hirabib}

\begin{thebibliography}{34}%
\makeatletter
\providecommand \@ifxundefined [1]{%
 \@ifx{#1\undefined}
}%
\providecommand \@ifnum [1]{%
 \ifnum #1\expandafter \@firstoftwo
 \else \expandafter \@secondoftwo
 \fi
}%
\providecommand \@ifx [1]{%
 \ifx #1\expandafter \@firstoftwo
 \else \expandafter \@secondoftwo
 \fi
}%
\providecommand \natexlab [1]{#1}%
\providecommand \enquote  [1]{``#1''}%
\providecommand \bibnamefont  [1]{#1}%
\providecommand \bibfnamefont [1]{#1}%
\providecommand \citenamefont [1]{#1}%
\providecommand \href@noop [0]{\@secondoftwo}%
\providecommand \href [0]{\begingroup \@sanitize@url \@href}%
\providecommand \@href[1]{\@@startlink{#1}\@@href}%
\providecommand \@@href[1]{\endgroup#1\@@endlink}%
\providecommand \@sanitize@url [0]{\catcode `\\12\catcode `\$12\catcode
  `\&12\catcode `\#12\catcode `\^12\catcode `\_12\catcode `\%12\relax}%
\providecommand \@@startlink[1]{}%
\providecommand \@@endlink[0]{}%
\providecommand \url  [0]{\begingroup\@sanitize@url \@url }%
\providecommand \@url [1]{\endgroup\@href {#1}{\urlprefix }}%
\providecommand \urlprefix  [0]{URL }%
\providecommand \Eprint [0]{\href }%
\providecommand \doibase [0]{https://doi.org/}%
\providecommand \selectlanguage [0]{\@gobble}%
\providecommand \bibinfo  [0]{\@secondoftwo}%
\providecommand \bibfield  [0]{\@secondoftwo}%
\providecommand \translation [1]{[#1]}%
\providecommand \BibitemOpen [0]{}%
\providecommand \bibitemStop [0]{}%
\providecommand \bibitemNoStop [0]{.\EOS\space}%
\providecommand \EOS [0]{\spacefactor3000\relax}%
\providecommand \BibitemShut  [1]{\csname bibitem#1\endcsname}%
\let\auto@bib@innerbib\@empty
\bibitem [{\citenamefont {McCann}\ and\ \citenamefont
  {Koshino}(2013)}]{Mcan13}%
  \BibitemOpen
  \bibfield  {author} {\bibinfo {author} {\bibfnamefont {E.}~\bibnamefont
  {McCann}}\ and\ \bibinfo {author} {\bibfnamefont {M.}~\bibnamefont
  {Koshino}},\ }\bibfield  {title} {\bibinfo {title} {The electronic properties
  of bilayer graphene},\ }\href {https://doi.org/10.1088/0034-4885/76/5/056503}
  {\bibfield  {journal} {\bibinfo  {journal} {Reports on Progress in Physics}\
  }\textbf {\bibinfo {volume} {76}},\ \bibinfo {pages} {056503} (\bibinfo
  {year} {2013})}\BibitemShut {NoStop}%
\bibitem [{\citenamefont {Zhang}\ \emph {et~al.}(2013)\citenamefont {Zhang},
  \citenamefont {MacDonald},\ and\ \citenamefont {Mele}}]{Zhang13}%
  \BibitemOpen
  \bibfield  {author} {\bibinfo {author} {\bibfnamefont {F.}~\bibnamefont
  {Zhang}}, \bibinfo {author} {\bibfnamefont {A.~H.}\ \bibnamefont
  {MacDonald}},\ and\ \bibinfo {author} {\bibfnamefont {E.~J.}\ \bibnamefont
  {Mele}},\ }\bibfield  {title} {\bibinfo {title} {Valley chern numbers and
  boundary modes in gapped bilayer graphene},\ }\href
  {https://doi.org/10.1073/pnas.1308853110} {\bibfield  {journal} {\bibinfo
  {journal} {Proceedings of the National Academy of Sciences}\ }\textbf
  {\bibinfo {volume} {110}},\ \bibinfo {pages} {10546} (\bibinfo {year}
  {2013})}\BibitemShut {NoStop}%
\bibitem [{\citenamefont {Rozhkov}\ \emph {et~al.}(2016)\citenamefont
  {Rozhkov}, \citenamefont {Sboychakov}, \citenamefont {Rakhmanov},\ and\
  \citenamefont {Nori}}]{rozhkov16}%
  \BibitemOpen
  \bibfield  {author} {\bibinfo {author} {\bibfnamefont {A.}~\bibnamefont
  {Rozhkov}}, \bibinfo {author} {\bibfnamefont {A.}~\bibnamefont {Sboychakov}},
  \bibinfo {author} {\bibfnamefont {A.}~\bibnamefont {Rakhmanov}},\ and\
  \bibinfo {author} {\bibfnamefont {F.}~\bibnamefont {Nori}},\ }\bibfield
  {title} {\bibinfo {title} {Electronic properties of graphene-based bilayer
  systems},\ }\href
  {https://doi.org/https://doi.org/10.1016/j.physrep.2016.07.003} {\bibfield
  {journal} {\bibinfo  {journal} {Physics Reports}\ }\textbf {\bibinfo {volume}
  {648}},\ \bibinfo {pages} {1} (\bibinfo {year} {2016})},\ \bibinfo {note}
  {electronic properties of graphene-based bilayer systems}\BibitemShut
  {NoStop}%
\bibitem [{\citenamefont {Overweg}\ \emph {et~al.}(2018)\citenamefont
  {Overweg}, \citenamefont {Knothe}, \citenamefont {Fabian}, \citenamefont
  {Linhart}, \citenamefont {Rickhaus}, \citenamefont {Wernli}, \citenamefont
  {Watanabe}, \citenamefont {Taniguchi}, \citenamefont {S\'anchez},
  \citenamefont {Burgd\"orfer}, \citenamefont {Libisch}, \citenamefont
  {Fal{'k}o}, \citenamefont {Ensslin},\ and\ \citenamefont {Ihn}}]{Over18}%
  \BibitemOpen
  \bibfield  {author} {\bibinfo {author} {\bibfnamefont {H.}~\bibnamefont
  {Overweg}}, \bibinfo {author} {\bibfnamefont {A.}~\bibnamefont {Knothe}},
  \bibinfo {author} {\bibfnamefont {T.}~\bibnamefont {Fabian}}, \bibinfo
  {author} {\bibfnamefont {L.}~\bibnamefont {Linhart}}, \bibinfo {author}
  {\bibfnamefont {P.}~\bibnamefont {Rickhaus}}, \bibinfo {author}
  {\bibfnamefont {L.}~\bibnamefont {Wernli}}, \bibinfo {author} {\bibfnamefont
  {K.}~\bibnamefont {Watanabe}}, \bibinfo {author} {\bibfnamefont
  {T.}~\bibnamefont {Taniguchi}}, \bibinfo {author} {\bibfnamefont
  {D.}~\bibnamefont {S\'anchez}}, \bibinfo {author} {\bibfnamefont
  {J.}~\bibnamefont {Burgd\"orfer}}, \bibinfo {author} {\bibfnamefont
  {F.}~\bibnamefont {Libisch}}, \bibinfo {author} {\bibfnamefont {V.~I.}\
  \bibnamefont {Fal{'k}o}}, \bibinfo {author} {\bibfnamefont {K.}~\bibnamefont
  {Ensslin}},\ and\ \bibinfo {author} {\bibfnamefont {T.}~\bibnamefont {Ihn}},\
  }\bibfield  {title} {\bibinfo {title} {Topologically nontrivial valley states
  in bilayer graphene quantum point contacts},\ }\href
  {https://doi.org/10.1103/PhysRevLett.121.257702} {\bibfield  {journal}
  {\bibinfo  {journal} {Phys. Rev. Lett.}\ }\textbf {\bibinfo {volume} {121}},\
  \bibinfo {pages} {257702} (\bibinfo {year} {2018})}\BibitemShut {NoStop}%
\bibitem [{\citenamefont {Kraft}\ \emph {et~al.}(2018)\citenamefont {Kraft},
  \citenamefont {Krainov}, \citenamefont {Gall}, \citenamefont {Dmitriev},
  \citenamefont {Krupke}, \citenamefont {Gornyi},\ and\ \citenamefont
  {Danneau}}]{Kraf18}%
  \BibitemOpen
  \bibfield  {author} {\bibinfo {author} {\bibfnamefont {R.}~\bibnamefont
  {Kraft}}, \bibinfo {author} {\bibfnamefont {I.~V.}\ \bibnamefont {Krainov}},
  \bibinfo {author} {\bibfnamefont {V.}~\bibnamefont {Gall}}, \bibinfo {author}
  {\bibfnamefont {A.~P.}\ \bibnamefont {Dmitriev}}, \bibinfo {author}
  {\bibfnamefont {R.}~\bibnamefont {Krupke}}, \bibinfo {author} {\bibfnamefont
  {I.~V.}\ \bibnamefont {Gornyi}},\ and\ \bibinfo {author} {\bibfnamefont
  {R.}~\bibnamefont {Danneau}},\ }\bibfield  {title} {\bibinfo {title} {Valley
  subband splitting in bilayer graphene quantum point contacts},\ }\href
  {https://doi.org/10.1103/PhysRevLett.121.257703} {\bibfield  {journal}
  {\bibinfo  {journal} {Phys. Rev. Lett.}\ }\textbf {\bibinfo {volume} {121}},\
  \bibinfo {pages} {257703} (\bibinfo {year} {2018})}\BibitemShut {NoStop}%
\bibitem [{\citenamefont {Eich}\ \emph {et~al.}(2018)\citenamefont {Eich},
  \citenamefont {Herman}, \citenamefont {Pisoni}, \citenamefont {Overweg},
  \citenamefont {Kurzmann}, \citenamefont {Lee}, \citenamefont {Rickhaus},
  \citenamefont {Watanabe}, \citenamefont {Taniguchi}, \citenamefont {Sigrist},
  \citenamefont {Ihn},\ and\ \citenamefont {Ensslin}}]{Eich18}%
  \BibitemOpen
  \bibfield  {author} {\bibinfo {author} {\bibfnamefont {M.}~\bibnamefont
  {Eich}}, \bibinfo {author} {\bibfnamefont {F.}~\bibnamefont {Herman}},
  \bibinfo {author} {\bibfnamefont {R.}~\bibnamefont {Pisoni}}, \bibinfo
  {author} {\bibfnamefont {H.}~\bibnamefont {Overweg}}, \bibinfo {author}
  {\bibfnamefont {A.}~\bibnamefont {Kurzmann}}, \bibinfo {author}
  {\bibfnamefont {Y.}~\bibnamefont {Lee}}, \bibinfo {author} {\bibfnamefont
  {P.}~\bibnamefont {Rickhaus}}, \bibinfo {author} {\bibfnamefont
  {K.}~\bibnamefont {Watanabe}}, \bibinfo {author} {\bibfnamefont
  {T.}~\bibnamefont {Taniguchi}}, \bibinfo {author} {\bibfnamefont
  {M.}~\bibnamefont {Sigrist}}, \bibinfo {author} {\bibfnamefont
  {T.}~\bibnamefont {Ihn}},\ and\ \bibinfo {author} {\bibfnamefont
  {K.}~\bibnamefont {Ensslin}},\ }\bibfield  {title} {\bibinfo {title} {Spin
  and valley states in gate-defined bilayer graphene quantum dots},\ }\href
  {https://doi.org/10.1103/PhysRevX.8.031023} {\bibfield  {journal} {\bibinfo
  {journal} {Phys. Rev. X}\ }\textbf {\bibinfo {volume} {8}},\ \bibinfo {pages}
  {031023} (\bibinfo {year} {2018})}\BibitemShut {NoStop}%
\bibitem [{\citenamefont {Kurzmann}\ \emph {et~al.}(2019)\citenamefont
  {Kurzmann}, \citenamefont {Overweg}, \citenamefont {Eich}, \citenamefont
  {Pally}, \citenamefont {Rickhaus}, \citenamefont {Pisoni}, \citenamefont
  {Lee}, \citenamefont {Watanabe}, \citenamefont {Taniguchi}, \citenamefont
  {Ihn},\ and\ \citenamefont {Ensslin}}]{Kurzmann19}%
  \BibitemOpen
  \bibfield  {author} {\bibinfo {author} {\bibfnamefont {A.}~\bibnamefont
  {Kurzmann}}, \bibinfo {author} {\bibfnamefont {H.}~\bibnamefont {Overweg}},
  \bibinfo {author} {\bibfnamefont {M.}~\bibnamefont {Eich}}, \bibinfo {author}
  {\bibfnamefont {A.}~\bibnamefont {Pally}}, \bibinfo {author} {\bibfnamefont
  {P.}~\bibnamefont {Rickhaus}}, \bibinfo {author} {\bibfnamefont
  {R.}~\bibnamefont {Pisoni}}, \bibinfo {author} {\bibfnamefont
  {Y.}~\bibnamefont {Lee}}, \bibinfo {author} {\bibfnamefont {K.}~\bibnamefont
  {Watanabe}}, \bibinfo {author} {\bibfnamefont {T.}~\bibnamefont {Taniguchi}},
  \bibinfo {author} {\bibfnamefont {T.}~\bibnamefont {Ihn}},\ and\ \bibinfo
  {author} {\bibfnamefont {K.}~\bibnamefont {Ensslin}},\ }\bibfield  {title}
  {\bibinfo {title} {Charge detection in gate-defined bilayer graphene quantum
  dots},\ }\href {https://doi.org/10.1021/acs.nanolett.9b01617} {\bibfield
  {journal} {\bibinfo  {journal} {Nano Letters}\ }\textbf {\bibinfo {volume}
  {19}},\ \bibinfo {pages} {5216} (\bibinfo {year} {2019})}\BibitemShut
  {NoStop}%
\bibitem [{\citenamefont {Banszerus}\ \emph {et~al.}(2020)\citenamefont
  {Banszerus}, \citenamefont {Rothstein}, \citenamefont {Fabian}, \citenamefont
  {M{\"o}ller}, \citenamefont {Icking}, \citenamefont {Trellenkamp},
  \citenamefont {Lentz}, \citenamefont {Neumaier}, \citenamefont {Watanabe},
  \citenamefont {Taniguchi}, \citenamefont {Libisch}, \citenamefont {Volk},\
  and\ \citenamefont {Stampfer}}]{Banszerus20}%
  \BibitemOpen
  \bibfield  {author} {\bibinfo {author} {\bibfnamefont {L.}~\bibnamefont
  {Banszerus}}, \bibinfo {author} {\bibfnamefont {A.}~\bibnamefont
  {Rothstein}}, \bibinfo {author} {\bibfnamefont {T.}~\bibnamefont {Fabian}},
  \bibinfo {author} {\bibfnamefont {S.}~\bibnamefont {M{\"o}ller}}, \bibinfo
  {author} {\bibfnamefont {E.}~\bibnamefont {Icking}}, \bibinfo {author}
  {\bibfnamefont {S.}~\bibnamefont {Trellenkamp}}, \bibinfo {author}
  {\bibfnamefont {F.}~\bibnamefont {Lentz}}, \bibinfo {author} {\bibfnamefont
  {D.}~\bibnamefont {Neumaier}}, \bibinfo {author} {\bibfnamefont
  {K.}~\bibnamefont {Watanabe}}, \bibinfo {author} {\bibfnamefont
  {T.}~\bibnamefont {Taniguchi}}, \bibinfo {author} {\bibfnamefont
  {F.}~\bibnamefont {Libisch}}, \bibinfo {author} {\bibfnamefont
  {C.}~\bibnamefont {Volk}},\ and\ \bibinfo {author} {\bibfnamefont
  {C.}~\bibnamefont {Stampfer}},\ }\bibfield  {title} {\bibinfo {title}
  {Electron hole crossover in gate-controlled bilayer graphene quantum dots},\
  }\href {https://doi.org/10.1021/acs.nanolett.0c03227} {\bibfield  {journal}
  {\bibinfo  {journal} {Nano Letters}\ }\textbf {\bibinfo {volume} {20}},\
  \bibinfo {pages} {7709} (\bibinfo {year} {2020})}\BibitemShut {NoStop}%
\bibitem [{\citenamefont {Banszerus}\ \emph {et~al.}(2021)\citenamefont
  {Banszerus}, \citenamefont {Hecker}, \citenamefont {Icking}, \citenamefont
  {Trellenkamp}, \citenamefont {Lentz}, \citenamefont {Neumaier}, \citenamefont
  {Watanabe}, \citenamefont {Taniguchi}, \citenamefont {Volk},\ and\
  \citenamefont {Stampfer}}]{Banszerus21}%
  \BibitemOpen
  \bibfield  {author} {\bibinfo {author} {\bibfnamefont {L.}~\bibnamefont
  {Banszerus}}, \bibinfo {author} {\bibfnamefont {K.}~\bibnamefont {Hecker}},
  \bibinfo {author} {\bibfnamefont {E.}~\bibnamefont {Icking}}, \bibinfo
  {author} {\bibfnamefont {S.}~\bibnamefont {Trellenkamp}}, \bibinfo {author}
  {\bibfnamefont {F.}~\bibnamefont {Lentz}}, \bibinfo {author} {\bibfnamefont
  {D.}~\bibnamefont {Neumaier}}, \bibinfo {author} {\bibfnamefont
  {K.}~\bibnamefont {Watanabe}}, \bibinfo {author} {\bibfnamefont
  {T.}~\bibnamefont {Taniguchi}}, \bibinfo {author} {\bibfnamefont
  {C.}~\bibnamefont {Volk}},\ and\ \bibinfo {author} {\bibfnamefont
  {C.}~\bibnamefont {Stampfer}},\ }\bibfield  {title} {\bibinfo {title}
  {Pulsed-gate spectroscopy of single-electron spin states in bilayer graphene
  quantum dots},\ }\href {https://doi.org/10.1103/PhysRevB.103.L081404}
  {\bibfield  {journal} {\bibinfo  {journal} {Phys. Rev. B}\ }\textbf {\bibinfo
  {volume} {103}},\ \bibinfo {pages} {L081404} (\bibinfo {year}
  {2021})}\BibitemShut {NoStop}%
\bibitem [{\citenamefont {Banszerus}\ \emph {et~al.}(2023)\citenamefont
  {Banszerus}, \citenamefont {M{\"o}ller}, \citenamefont {Hecker},
  \citenamefont {Icking}, \citenamefont {Watanabe}, \citenamefont {Taniguchi},
  \citenamefont {Hassler}, \citenamefont {Volk},\ and\ \citenamefont
  {Stampfer}}]{Ban23}%
  \BibitemOpen
  \bibfield  {author} {\bibinfo {author} {\bibfnamefont {L.}~\bibnamefont
  {Banszerus}}, \bibinfo {author} {\bibfnamefont {S.}~\bibnamefont
  {M{\"o}ller}}, \bibinfo {author} {\bibfnamefont {K.}~\bibnamefont {Hecker}},
  \bibinfo {author} {\bibfnamefont {E.}~\bibnamefont {Icking}}, \bibinfo
  {author} {\bibfnamefont {K.}~\bibnamefont {Watanabe}}, \bibinfo {author}
  {\bibfnamefont {T.}~\bibnamefont {Taniguchi}}, \bibinfo {author}
  {\bibfnamefont {F.}~\bibnamefont {Hassler}}, \bibinfo {author} {\bibfnamefont
  {C.}~\bibnamefont {Volk}},\ and\ \bibinfo {author} {\bibfnamefont
  {C.}~\bibnamefont {Stampfer}},\ }\bibfield  {title} {\bibinfo {title}
  {Particle--hole symmetry protects spin-valley blockade in graphene quantum
  dots},\ }\bibfield  {journal} {\bibinfo  {journal} {Nature}\ }\href
  {https://doi.org/10.1038/s41586-023-05953-5} {10.1038/s41586-023-05953-5}
  (\bibinfo {year} {2023})\BibitemShut {NoStop}%
\bibitem [{\citenamefont {Meng}\ \emph {et~al.}(2012)\citenamefont {Meng},
  \citenamefont {Chu}, \citenamefont {Zhang}, \citenamefont {Yang},
  \citenamefont {Dou}, \citenamefont {Nie},\ and\ \citenamefont {He}}]{Men12}%
  \BibitemOpen
  \bibfield  {author} {\bibinfo {author} {\bibfnamefont {L.}~\bibnamefont
  {Meng}}, \bibinfo {author} {\bibfnamefont {Z.-D.}\ \bibnamefont {Chu}},
  \bibinfo {author} {\bibfnamefont {Y.}~\bibnamefont {Zhang}}, \bibinfo
  {author} {\bibfnamefont {J.-Y.}\ \bibnamefont {Yang}}, \bibinfo {author}
  {\bibfnamefont {R.-F.}\ \bibnamefont {Dou}}, \bibinfo {author} {\bibfnamefont
  {J.-C.}\ \bibnamefont {Nie}},\ and\ \bibinfo {author} {\bibfnamefont
  {L.}~\bibnamefont {He}},\ }\bibfield  {title} {\bibinfo {title} {Enhanced
  intervalley scattering of twisted bilayer graphene by periodic $ab$ stacked
  atoms},\ }\href {https://doi.org/10.1103/PhysRevB.85.235453} {\bibfield
  {journal} {\bibinfo  {journal} {Phys. Rev. B}\ }\textbf {\bibinfo {volume}
  {85}},\ \bibinfo {pages} {235453} (\bibinfo {year} {2012})}\BibitemShut
  {NoStop}%
\bibitem [{\citenamefont {Cleric\`o}\ \emph {et~al.}(2019)\citenamefont
  {Cleric\`o}, \citenamefont {Delgado-Notario}, \citenamefont
  {Saiz-Bret\'{\i}n}, \citenamefont {Malyshev}, \citenamefont {Meziani},
  \citenamefont {Hidalgo}, \citenamefont {M\'endez}, \citenamefont {Amado},
  \citenamefont {Dom\'{\i}nguez-Adame},\ and\ \citenamefont {Diez}}]{Cle19}%
  \BibitemOpen
  \bibfield  {author} {\bibinfo {author} {\bibfnamefont {V.}~\bibnamefont
  {Cleric\`o}}, \bibinfo {author} {\bibfnamefont {J.~A.}\ \bibnamefont
  {Delgado-Notario}}, \bibinfo {author} {\bibfnamefont {M.}~\bibnamefont
  {Saiz-Bret\'{\i}n}}, \bibinfo {author} {\bibfnamefont {A.~V.}\ \bibnamefont
  {Malyshev}}, \bibinfo {author} {\bibfnamefont {Y.~M.}\ \bibnamefont
  {Meziani}}, \bibinfo {author} {\bibfnamefont {P.}~\bibnamefont {Hidalgo}},
  \bibinfo {author} {\bibfnamefont {B.}~\bibnamefont {M\'endez}}, \bibinfo
  {author} {\bibfnamefont {M.}~\bibnamefont {Amado}}, \bibinfo {author}
  {\bibfnamefont {F.}~\bibnamefont {Dom\'{\i}nguez-Adame}},\ and\ \bibinfo
  {author} {\bibfnamefont {E.}~\bibnamefont {Diez}},\ }\bibfield  {title}
  {\bibinfo {title} {Quantum nanoconstrictions fabricated by cryo-etching in
  encapsulated graphene},\ }\href@noop {} {\bibfield  {journal} {\bibinfo
  {journal} {Sci. Rep.}\ }\textbf {\bibinfo {volume} {9}},\ \bibinfo {pages}
  {13572} (\bibinfo {year} {2019})}\BibitemShut {NoStop}%
\bibitem [{\citenamefont {Jin}\ \emph {et~al.}(2021)\citenamefont {Jin},
  \citenamefont {Zong}, \citenamefont {Chen}, \citenamefont {Tian},
  \citenamefont {Qiu}, \citenamefont {Liu}, \citenamefont {Zheng},
  \citenamefont {Xi}, \citenamefont {Gao}, \citenamefont {Wang},\ and\
  \citenamefont {Zhang}}]{Jin21}%
  \BibitemOpen
  \bibfield  {author} {\bibinfo {author} {\bibfnamefont {S.}~\bibnamefont
  {Jin}}, \bibinfo {author} {\bibfnamefont {J.}~\bibnamefont {Zong}}, \bibinfo
  {author} {\bibfnamefont {W.}~\bibnamefont {Chen}}, \bibinfo {author}
  {\bibfnamefont {Q.}~\bibnamefont {Tian}}, \bibinfo {author} {\bibfnamefont
  {X.}~\bibnamefont {Qiu}}, \bibinfo {author} {\bibfnamefont {G.}~\bibnamefont
  {Liu}}, \bibinfo {author} {\bibfnamefont {H.}~\bibnamefont {Zheng}}, \bibinfo
  {author} {\bibfnamefont {X.}~\bibnamefont {Xi}}, \bibinfo {author}
  {\bibfnamefont {L.}~\bibnamefont {Gao}}, \bibinfo {author} {\bibfnamefont
  {C.}~\bibnamefont {Wang}},\ and\ \bibinfo {author} {\bibfnamefont
  {Y.}~\bibnamefont {Zhang}},\ }\bibfield  {title} {\bibinfo {title} {Epitaxial
  growth of uniform single-layer and bilayer graphene with assistance of
  nitrogen plasma},\ }\bibfield  {journal} {\bibinfo  {journal}
  {Nanomaterials}\ }\textbf {\bibinfo {volume} {11}},\ \href
  {https://doi.org/10.3390/nano11123217} {10.3390/nano11123217} (\bibinfo
  {year} {2021})\BibitemShut {NoStop}%
\bibitem [{\citenamefont {Castro}\ \emph {et~al.}(2007)\citenamefont {Castro},
  \citenamefont {Novoselov}, \citenamefont {Morozov}, \citenamefont {Peres},
  \citenamefont {dos Santos}, \citenamefont {Nilsson}, \citenamefont {Guinea},
  \citenamefont {Geim},\ and\ \citenamefont {Neto}}]{Cas07}%
  \BibitemOpen
  \bibfield  {author} {\bibinfo {author} {\bibfnamefont {E.~V.}\ \bibnamefont
  {Castro}}, \bibinfo {author} {\bibfnamefont {K.~S.}\ \bibnamefont
  {Novoselov}}, \bibinfo {author} {\bibfnamefont {S.~V.}\ \bibnamefont
  {Morozov}}, \bibinfo {author} {\bibfnamefont {N.~M.~R.}\ \bibnamefont
  {Peres}}, \bibinfo {author} {\bibfnamefont {J.~M. B.~L.}\ \bibnamefont {dos
  Santos}}, \bibinfo {author} {\bibfnamefont {J.}~\bibnamefont {Nilsson}},
  \bibinfo {author} {\bibfnamefont {F.}~\bibnamefont {Guinea}}, \bibinfo
  {author} {\bibfnamefont {A.~K.}\ \bibnamefont {Geim}},\ and\ \bibinfo
  {author} {\bibfnamefont {A.~H.~C.}\ \bibnamefont {Neto}},\ }\bibfield
  {title} {\bibinfo {title} {Biased bilayer graphene: Semiconductor with a gap
  tunable by the electric field effect},\ }\href
  {https://doi.org/10.1103/PhysRevLett.99.216802} {\bibfield  {journal}
  {\bibinfo  {journal} {Phys. Rev. Lett.}\ }\textbf {\bibinfo {volume} {99}},\
  \bibinfo {pages} {216802} (\bibinfo {year} {2007})}\BibitemShut {NoStop}%
\bibitem [{\citenamefont {Ju}\ \emph {et~al.}(2015)\citenamefont {Ju},
  \citenamefont {Shi}, \citenamefont {Nair}, \citenamefont {Lv}, \citenamefont
  {Jin}, \citenamefont {Velasco}, \citenamefont {Ojeda-Aristizabal},
  \citenamefont {Bechtel}, \citenamefont {Martin}, \citenamefont {Zettl},
  \citenamefont {Analytis},\ and\ \citenamefont {Wang}}]{Lon15}%
  \BibitemOpen
  \bibfield  {author} {\bibinfo {author} {\bibfnamefont {L.}~\bibnamefont
  {Ju}}, \bibinfo {author} {\bibfnamefont {Z.}~\bibnamefont {Shi}}, \bibinfo
  {author} {\bibfnamefont {N.}~\bibnamefont {Nair}}, \bibinfo {author}
  {\bibfnamefont {Y.}~\bibnamefont {Lv}}, \bibinfo {author} {\bibfnamefont
  {C.}~\bibnamefont {Jin}}, \bibinfo {author} {\bibfnamefont {J.}~\bibnamefont
  {Velasco}}, \bibinfo {author} {\bibfnamefont {C.}~\bibnamefont
  {Ojeda-Aristizabal}}, \bibinfo {author} {\bibfnamefont {H.~A.}\ \bibnamefont
  {Bechtel}}, \bibinfo {author} {\bibfnamefont {M.~C.}\ \bibnamefont {Martin}},
  \bibinfo {author} {\bibfnamefont {A.}~\bibnamefont {Zettl}}, \bibinfo
  {author} {\bibfnamefont {J.}~\bibnamefont {Analytis}},\ and\ \bibinfo
  {author} {\bibfnamefont {F.}~\bibnamefont {Wang}},\ }\bibfield  {title}
  {\bibinfo {title} {Topological valley transport at bilayer graphene domain
  walls},\ }\href {https://doi.org/10.1038/nature14364} {\bibfield  {journal}
  {\bibinfo  {journal} {Nature}\ }\textbf {\bibinfo {volume} {520}},\ \bibinfo
  {pages} {650} (\bibinfo {year} {2015})}\BibitemShut {NoStop}%
\bibitem [{\citenamefont {Sui}\ \emph {et~al.}(2015)\citenamefont {Sui},
  \citenamefont {Chen}, \citenamefont {Ma}, \citenamefont {Shan}, \citenamefont
  {Tian}, \citenamefont {Watanabe}, \citenamefont {Taniguchi}, \citenamefont
  {Jin}, \citenamefont {Yao}, \citenamefont {Xiao},\ and\ \citenamefont
  {Zhang}}]{Men15}%
  \BibitemOpen
  \bibfield  {author} {\bibinfo {author} {\bibfnamefont {M.}~\bibnamefont
  {Sui}}, \bibinfo {author} {\bibfnamefont {G.}~\bibnamefont {Chen}}, \bibinfo
  {author} {\bibfnamefont {L.}~\bibnamefont {Ma}}, \bibinfo {author}
  {\bibfnamefont {W.-Y.}\ \bibnamefont {Shan}}, \bibinfo {author}
  {\bibfnamefont {D.}~\bibnamefont {Tian}}, \bibinfo {author} {\bibfnamefont
  {K.}~\bibnamefont {Watanabe}}, \bibinfo {author} {\bibfnamefont
  {T.}~\bibnamefont {Taniguchi}}, \bibinfo {author} {\bibfnamefont
  {X.}~\bibnamefont {Jin}}, \bibinfo {author} {\bibfnamefont {W.}~\bibnamefont
  {Yao}}, \bibinfo {author} {\bibfnamefont {D.}~\bibnamefont {Xiao}},\ and\
  \bibinfo {author} {\bibfnamefont {Y.}~\bibnamefont {Zhang}},\ }\bibfield
  {title} {\bibinfo {title} {Gate-tunable topological valley transport in
  bilayer graphene},\ }\href {https://doi.org/10.1038/nphys3485} {\bibfield
  {journal} {\bibinfo  {journal} {Nature Physics}\ }\textbf {\bibinfo {volume}
  {11}},\ \bibinfo {pages} {1027} (\bibinfo {year} {2015})}\BibitemShut
  {NoStop}%
\bibitem [{\citenamefont {Li}\ \emph {et~al.}(2016)\citenamefont {Li},
  \citenamefont {Wang}, \citenamefont {McFaul}, \citenamefont {Zern},
  \citenamefont {Ren}, \citenamefont {Watanabe}, \citenamefont {Taniguchi},
  \citenamefont {Qiao},\ and\ \citenamefont {Zhu}}]{Li16}%
  \BibitemOpen
  \bibfield  {author} {\bibinfo {author} {\bibfnamefont {J.}~\bibnamefont
  {Li}}, \bibinfo {author} {\bibfnamefont {K.}~\bibnamefont {Wang}}, \bibinfo
  {author} {\bibfnamefont {K.~J.}\ \bibnamefont {McFaul}}, \bibinfo {author}
  {\bibfnamefont {Z.}~\bibnamefont {Zern}}, \bibinfo {author} {\bibfnamefont
  {Y.}~\bibnamefont {Ren}}, \bibinfo {author} {\bibfnamefont {K.}~\bibnamefont
  {Watanabe}}, \bibinfo {author} {\bibfnamefont {T.}~\bibnamefont {Taniguchi}},
  \bibinfo {author} {\bibfnamefont {Z.}~\bibnamefont {Qiao}},\ and\ \bibinfo
  {author} {\bibfnamefont {J.}~\bibnamefont {Zhu}},\ }\bibfield  {title}
  {\bibinfo {title} {Gate-controlled topological conducting channels in bilayer
  graphene},\ }\href {https://doi.org/10.1038/nnano.2016.158} {\bibfield
  {journal} {\bibinfo  {journal} {Nature Nanotechnology}\ }\textbf {\bibinfo
  {volume} {11}},\ \bibinfo {pages} {1060} (\bibinfo {year}
  {2016})}\BibitemShut {NoStop}%
\bibitem [{\citenamefont {Chen}\ \emph {et~al.}(2020)\citenamefont {Chen},
  \citenamefont {Zhou}, \citenamefont {Liu}, \citenamefont {Qiao},
  \citenamefont {Oezyilmaz},\ and\ \citenamefont {Martin}}]{Chen20}%
  \BibitemOpen
  \bibfield  {author} {\bibinfo {author} {\bibfnamefont {H.}~\bibnamefont
  {Chen}}, \bibinfo {author} {\bibfnamefont {P.}~\bibnamefont {Zhou}}, \bibinfo
  {author} {\bibfnamefont {J.}~\bibnamefont {Liu}}, \bibinfo {author}
  {\bibfnamefont {J.}~\bibnamefont {Qiao}}, \bibinfo {author} {\bibfnamefont
  {B.}~\bibnamefont {Oezyilmaz}},\ and\ \bibinfo {author} {\bibfnamefont
  {J.}~\bibnamefont {Martin}},\ }\bibfield  {title} {\bibinfo {title} {Gate
  controlled valley polarizer in bilayer graphene},\ }\href
  {https://doi.org/10.1038/s41467-020-15117-y} {\bibfield  {journal} {\bibinfo
  {journal} {Nature Communications}\ }\textbf {\bibinfo {volume} {11}},\
  \bibinfo {pages} {1202} (\bibinfo {year} {2020})}\BibitemShut {NoStop}%
\bibitem [{\citenamefont {Ryu}\ \emph {et~al.}(2022)\citenamefont {Ryu},
  \citenamefont {L\'opez},\ and\ \citenamefont {Serra}}]{Ryu22}%
  \BibitemOpen
  \bibfield  {author} {\bibinfo {author} {\bibfnamefont {S.}~\bibnamefont
  {Ryu}}, \bibinfo {author} {\bibfnamefont {R.}~\bibnamefont {L\'opez}},\ and\
  \bibinfo {author} {\bibfnamefont {L.}~\bibnamefont {Serra}},\ }\bibfield
  {title} {\bibinfo {title} {Conductance of electrostatic wire junctions in
  bilayer graphene},\ }\href {https://doi.org/10.1103/PhysRevB.106.035424}
  {\bibfield  {journal} {\bibinfo  {journal} {Phys. Rev. B}\ }\textbf {\bibinfo
  {volume} {106}},\ \bibinfo {pages} {035424} (\bibinfo {year}
  {2022})}\BibitemShut {NoStop}%
\bibitem [{\citenamefont {Ihn}(2009)}]{Ihn}%
  \BibitemOpen
  \bibfield  {author} {\bibinfo {author} {\bibfnamefont {T.}~\bibnamefont
  {Ihn}},\ }\href@noop {} {\emph {\bibinfo {title} {Semiconductor
  Nanostructures: Quantum states and electronic transport}}}\ (\bibinfo
  {publisher} {Oxford University Press},\ \bibinfo {year} {2009})\BibitemShut
  {NoStop}%
\bibitem [{\citenamefont {Pereira}\ \emph {et~al.}(2007)\citenamefont
  {Pereira}, \citenamefont {Vasilopoulos},\ and\ \citenamefont
  {Peeters}}]{Pereira07}%
  \BibitemOpen
  \bibfield  {author} {\bibinfo {author} {\bibfnamefont {J.~M.}\ \bibnamefont
  {Pereira}}, \bibinfo {author} {\bibfnamefont {P.}~\bibnamefont
  {Vasilopoulos}},\ and\ \bibinfo {author} {\bibfnamefont {F.~M.}\ \bibnamefont
  {Peeters}},\ }\bibfield  {title} {\bibinfo {title} {Tunable quantum dots in
  bilayer graphene},\ }\href {https://doi.org/10.1021/nl062967s} {\bibfield
  {journal} {\bibinfo  {journal} {Nano Letters}\ }\textbf {\bibinfo {volume}
  {7}},\ \bibinfo {pages} {946} (\bibinfo {year} {2007})}\BibitemShut {NoStop}%
\bibitem [{\citenamefont {Recher}\ \emph {et~al.}(2009)\citenamefont {Recher},
  \citenamefont {Nilsson}, \citenamefont {Burkard},\ and\ \citenamefont
  {Trauzettel}}]{Recher09}%
  \BibitemOpen
  \bibfield  {author} {\bibinfo {author} {\bibfnamefont {P.}~\bibnamefont
  {Recher}}, \bibinfo {author} {\bibfnamefont {J.}~\bibnamefont {Nilsson}},
  \bibinfo {author} {\bibfnamefont {G.}~\bibnamefont {Burkard}},\ and\ \bibinfo
  {author} {\bibfnamefont {B.}~\bibnamefont {Trauzettel}},\ }\bibfield  {title}
  {\bibinfo {title} {Bound states and magnetic field induced valley splitting
  in gate-tunable graphene quantum dots},\ }\href
  {https://doi.org/10.1103/PhysRevB.79.085407} {\bibfield  {journal} {\bibinfo
  {journal} {Phys. Rev. B}\ }\textbf {\bibinfo {volume} {79}},\ \bibinfo
  {pages} {085407} (\bibinfo {year} {2009})}\BibitemShut {NoStop}%
\bibitem [{\citenamefont {Zarenia}\ \emph {et~al.}(2009)\citenamefont
  {Zarenia}, \citenamefont {Pereira}, \citenamefont {Peeters},\ and\
  \citenamefont {Farias}}]{Zarenia09}%
  \BibitemOpen
  \bibfield  {author} {\bibinfo {author} {\bibfnamefont {M.}~\bibnamefont
  {Zarenia}}, \bibinfo {author} {\bibfnamefont {J.~M.}\ \bibnamefont
  {Pereira}}, \bibinfo {author} {\bibfnamefont {F.~M.}\ \bibnamefont
  {Peeters}},\ and\ \bibinfo {author} {\bibfnamefont {G.~A.}\ \bibnamefont
  {Farias}},\ }\bibfield  {title} {\bibinfo {title} {Electrostatically confined
  quantum rings in bilayer graphene},\ }\href
  {https://doi.org/10.1021/nl902302m} {\bibfield  {journal} {\bibinfo
  {journal} {Nano Letters}\ }\textbf {\bibinfo {volume} {9}},\ \bibinfo {pages}
  {4088} (\bibinfo {year} {2009})}\BibitemShut {NoStop}%
\bibitem [{\citenamefont {Pereira}\ \emph {et~al.}(2009)\citenamefont
  {Pereira}, \citenamefont {Peeters}, \citenamefont {Vasilopoulos},
  \citenamefont {Costa~Filho},\ and\ \citenamefont {Farias}}]{Pereira09}%
  \BibitemOpen
  \bibfield  {author} {\bibinfo {author} {\bibfnamefont {J.~M.}\ \bibnamefont
  {Pereira}}, \bibinfo {author} {\bibfnamefont {F.~M.}\ \bibnamefont
  {Peeters}}, \bibinfo {author} {\bibfnamefont {P.}~\bibnamefont
  {Vasilopoulos}}, \bibinfo {author} {\bibfnamefont {R.~N.}\ \bibnamefont
  {Costa~Filho}},\ and\ \bibinfo {author} {\bibfnamefont {G.~A.}\ \bibnamefont
  {Farias}},\ }\bibfield  {title} {\bibinfo {title} {Landau levels in graphene
  bilayer quantum dots},\ }\href {https://doi.org/10.1103/PhysRevB.79.195403}
  {\bibfield  {journal} {\bibinfo  {journal} {Phys. Rev. B}\ }\textbf {\bibinfo
  {volume} {79}},\ \bibinfo {pages} {195403} (\bibinfo {year}
  {2009})}\BibitemShut {NoStop}%
\bibitem [{\citenamefont {Zarenia}\ \emph
  {et~al.}(2010{\natexlab{a}})\citenamefont {Zarenia}, \citenamefont {Pereira},
  \citenamefont {Chaves}, \citenamefont {Peeters},\ and\ \citenamefont
  {Farias}}]{Zarenia10}%
  \BibitemOpen
  \bibfield  {author} {\bibinfo {author} {\bibfnamefont {M.}~\bibnamefont
  {Zarenia}}, \bibinfo {author} {\bibfnamefont {J.~M.}\ \bibnamefont
  {Pereira}}, \bibinfo {author} {\bibfnamefont {A.}~\bibnamefont {Chaves}},
  \bibinfo {author} {\bibfnamefont {F.~M.}\ \bibnamefont {Peeters}},\ and\
  \bibinfo {author} {\bibfnamefont {G.~A.}\ \bibnamefont {Farias}},\ }\bibfield
   {title} {\bibinfo {title} {Simplified model for the energy levels of quantum
  rings in single layer and bilayer graphene},\ }\href
  {https://doi.org/10.1103/PhysRevB.81.045431} {\bibfield  {journal} {\bibinfo
  {journal} {Phys. Rev. B}\ }\textbf {\bibinfo {volume} {81}},\ \bibinfo
  {pages} {045431} (\bibinfo {year} {2010}{\natexlab{a}})}\BibitemShut
  {NoStop}%
\bibitem [{\citenamefont {Zarenia}\ \emph
  {et~al.}(2010{\natexlab{b}})\citenamefont {Zarenia}, \citenamefont {Pereira},
  \citenamefont {Chaves}, \citenamefont {Peeters},\ and\ \citenamefont
  {Farias}}]{Zarenia10b}%
  \BibitemOpen
  \bibfield  {author} {\bibinfo {author} {\bibfnamefont {M.}~\bibnamefont
  {Zarenia}}, \bibinfo {author} {\bibfnamefont {J.~M.}\ \bibnamefont
  {Pereira}}, \bibinfo {author} {\bibfnamefont {A.}~\bibnamefont {Chaves}},
  \bibinfo {author} {\bibfnamefont {F.~M.}\ \bibnamefont {Peeters}},\ and\
  \bibinfo {author} {\bibfnamefont {G.~A.}\ \bibnamefont {Farias}},\ }\bibfield
   {title} {\bibinfo {title} {Erratum: Simplified model for the energy levels
  of quantum rings in single layer and bilayer graphene [{P}hys. {R}ev. {B} 81,
  045431 (2010)]},\ }\href {https://doi.org/10.1103/PhysRevB.82.119906}
  {\bibfield  {journal} {\bibinfo  {journal} {Phys. Rev. B}\ }\textbf {\bibinfo
  {volume} {82}},\ \bibinfo {pages} {119906(E)} (\bibinfo {year}
  {2010}{\natexlab{b}})}\BibitemShut {NoStop}%
\bibitem [{\citenamefont {{da Costa}}\ \emph {et~al.}(2014)\citenamefont {{da
  Costa}}, \citenamefont {Zarenia}, \citenamefont {Chaves}, \citenamefont
  {Farias},\ and\ \citenamefont {Peeters}}]{daCosta14}%
  \BibitemOpen
  \bibfield  {author} {\bibinfo {author} {\bibfnamefont {D.}~\bibnamefont {{da
  Costa}}}, \bibinfo {author} {\bibfnamefont {M.}~\bibnamefont {Zarenia}},
  \bibinfo {author} {\bibfnamefont {A.}~\bibnamefont {Chaves}}, \bibinfo
  {author} {\bibfnamefont {G.}~\bibnamefont {Farias}},\ and\ \bibinfo {author}
  {\bibfnamefont {F.}~\bibnamefont {Peeters}},\ }\bibfield  {title} {\bibinfo
  {title} {Analytical study of the energy levels in bilayer graphene quantum
  dots},\ }\href {https://doi.org/https://doi.org/10.1016/j.carbon.2014.06.078}
  {\bibfield  {journal} {\bibinfo  {journal} {Carbon}\ }\textbf {\bibinfo
  {volume} {78}},\ \bibinfo {pages} {392} (\bibinfo {year} {2014})}\BibitemShut
  {NoStop}%
\bibitem [{\citenamefont {Li}\ \emph {et~al.}(2010)\citenamefont {Li},
  \citenamefont {Morpurgo}, \citenamefont {B\"uttiker},\ and\ \citenamefont
  {Martin}}]{Li10}%
  \BibitemOpen
  \bibfield  {author} {\bibinfo {author} {\bibfnamefont {J.}~\bibnamefont
  {Li}}, \bibinfo {author} {\bibfnamefont {A.~F.}\ \bibnamefont {Morpurgo}},
  \bibinfo {author} {\bibfnamefont {M.}~\bibnamefont {B\"uttiker}},\ and\
  \bibinfo {author} {\bibfnamefont {I.}~\bibnamefont {Martin}},\ }\bibfield
  {title} {\bibinfo {title} {Marginality of bulk-edge correspondence for
  single-valley hamiltonians},\ }\href
  {https://doi.org/10.1103/PhysRevB.82.245404} {\bibfield  {journal} {\bibinfo
  {journal} {Phys. Rev. B}\ }\textbf {\bibinfo {volume} {82}},\ \bibinfo
  {pages} {245404} (\bibinfo {year} {2010})}\BibitemShut {NoStop}%
\bibitem [{\citenamefont {Martin}\ \emph {et~al.}(2008)\citenamefont {Martin},
  \citenamefont {Blanter},\ and\ \citenamefont {Morpurgo}}]{Mar08}%
  \BibitemOpen
  \bibfield  {author} {\bibinfo {author} {\bibfnamefont {I.}~\bibnamefont
  {Martin}}, \bibinfo {author} {\bibfnamefont {Y.~M.}\ \bibnamefont
  {Blanter}},\ and\ \bibinfo {author} {\bibfnamefont {A.~F.}\ \bibnamefont
  {Morpurgo}},\ }\bibfield  {title} {\bibinfo {title} {Topological confinement
  in bilayer graphene},\ }\href
  {https://doi.org/10.1103/PhysRevLett.100.036804} {\bibfield  {journal}
  {\bibinfo  {journal} {Phys. Rev. Lett.}\ }\textbf {\bibinfo {volume} {100}},\
  \bibinfo {pages} {036804} (\bibinfo {year} {2008})}\BibitemShut {NoStop}%
\bibitem [{\citenamefont {Zarenia}\ \emph {et~al.}(2011)\citenamefont
  {Zarenia}, \citenamefont {Pereira}, \citenamefont {Farias},\ and\
  \citenamefont {Peeters}}]{Zarenia11}%
  \BibitemOpen
  \bibfield  {author} {\bibinfo {author} {\bibfnamefont {M.}~\bibnamefont
  {Zarenia}}, \bibinfo {author} {\bibfnamefont {J.~M.}\ \bibnamefont
  {Pereira}}, \bibinfo {author} {\bibfnamefont {G.~A.}\ \bibnamefont
  {Farias}},\ and\ \bibinfo {author} {\bibfnamefont {F.~M.}\ \bibnamefont
  {Peeters}},\ }\bibfield  {title} {\bibinfo {title} {Chiral states in bilayer
  graphene: Magnetic field dependence and gap opening},\ }\href
  {https://doi.org/10.1103/PhysRevB.84.125451} {\bibfield  {journal} {\bibinfo
  {journal} {Phys. Rev. B}\ }\textbf {\bibinfo {volume} {84}},\ \bibinfo
  {pages} {125451} (\bibinfo {year} {2011})}\BibitemShut {NoStop}%
\bibitem [{\citenamefont {Xavier}\ \emph {et~al.}(2010)\citenamefont {Xavier},
  \citenamefont {Pereira}, \citenamefont {Chaves}, \citenamefont {Farias},\
  and\ \citenamefont {Peeters}}]{xavier10}%
  \BibitemOpen
  \bibfield  {author} {\bibinfo {author} {\bibfnamefont {L.~J.~P.}\
  \bibnamefont {Xavier}}, \bibinfo {author} {\bibfnamefont {J.~M.}\
  \bibnamefont {Pereira}}, \bibinfo {author} {\bibfnamefont {A.}~\bibnamefont
  {Chaves}}, \bibinfo {author} {\bibfnamefont {G.~A.}\ \bibnamefont {Farias}},\
  and\ \bibinfo {author} {\bibfnamefont {F.~M.}\ \bibnamefont {Peeters}},\
  }\bibfield  {title} {\bibinfo {title} {Topological confinement in graphene
  bilayer quantum rings},\ }\href {https://doi.org/10.1063/1.3431618}
  {\bibfield  {journal} {\bibinfo  {journal} {Applied Physics Letters}\
  }\textbf {\bibinfo {volume} {96}},\ \bibinfo {pages} {212108} (\bibinfo
  {year} {2010})}\BibitemShut {NoStop}%
\bibitem [{\citenamefont {Benchtaber}\ \emph
  {et~al.}(2021{\natexlab{a}})\citenamefont {Benchtaber}, \citenamefont
  {S\'anchez},\ and\ \citenamefont {Serra}}]{Ben21}%
  \BibitemOpen
  \bibfield  {author} {\bibinfo {author} {\bibfnamefont {N.}~\bibnamefont
  {Benchtaber}}, \bibinfo {author} {\bibfnamefont {D.}~\bibnamefont
  {S\'anchez}},\ and\ \bibinfo {author} {\bibfnamefont {L.}~\bibnamefont
  {Serra}},\ }\bibfield  {title} {\bibinfo {title} {Scattering of topological
  kink-antikink states in bilayer graphene structures},\ }\href
  {https://doi.org/10.1103/PhysRevB.104.155303} {\bibfield  {journal} {\bibinfo
   {journal} {Phys. Rev. B}\ }\textbf {\bibinfo {volume} {104}},\ \bibinfo
  {pages} {155303} (\bibinfo {year} {2021}{\natexlab{a}})}\BibitemShut
  {NoStop}%
\bibitem [{\citenamefont {Benchtaber}\ \emph
  {et~al.}(2021{\natexlab{b}})\citenamefont {Benchtaber}, \citenamefont
  {S{\'{a}}nchez},\ and\ \citenamefont {Serra}}]{Ben21b}%
  \BibitemOpen
  \bibfield  {author} {\bibinfo {author} {\bibfnamefont {N.}~\bibnamefont
  {Benchtaber}}, \bibinfo {author} {\bibfnamefont {D.}~\bibnamefont
  {S{\'{a}}nchez}},\ and\ \bibinfo {author} {\bibfnamefont {L.}~\bibnamefont
  {Serra}},\ }\bibfield  {title} {\bibinfo {title} {Geometry effects in
  topologically confined bilayer graphene loops},\ }\href
  {https://doi.org/10.1088/1367-2630/ac434d} {\bibfield  {journal} {\bibinfo
  {journal} {New Journal of Physics}\ }\textbf {\bibinfo {volume} {24}},\
  \bibinfo {pages} {013001} (\bibinfo {year} {2021}{\natexlab{b}})}\BibitemShut
  {NoStop}%
\bibitem [{\citenamefont {Osca}\ and\ \citenamefont {Serra}(2019)}]{Osca19}%
  \BibitemOpen
  \bibfield  {author} {\bibinfo {author} {\bibfnamefont {J.}~\bibnamefont
  {Osca}}\ and\ \bibinfo {author} {\bibfnamefont {L.}~\bibnamefont {Serra}},\
  }\bibfield  {title} {\bibinfo {title} {Complex band-structure analysis and
  topological physics of {M}ajorana nanowires},\ }\href
  {https://doi.org/10.1140/epjb/e2019-100011-2} {\bibfield  {journal} {\bibinfo
   {journal} {Eur. Phys. J. B}\ }\textbf {\bibinfo {volume} {92}},\ \bibinfo
  {pages} {101} (\bibinfo {year} {2019})}\BibitemShut {NoStop}%
\end{thebibliography}%

\end{document}